\documentclass[epj,nopacs]{svjour}
\usepackage{psfig}
\newcommand{\di}{\mathrm{d}}
\newcommand{\mrm}[1]{\mathrm{#1}}
\newcommand{\an}{\mbox{\boldmath$\alpha$}_N}
\begin{document}
\title{Soft hadron production in $pp$ interactions up to ISR energies}
\author{H. M\"uller
}                     
%
%
\institute{Institut f\"ur Kern- und Hadronenphysik,
Forschungszentrum Rossendorf, Postfach 510119, 01314 Dresden,
Germany}
\date{Received: date / Revised version: date}
%
\abstract{
  Soft hadron production is described  as a two-step process, where  the
  interaction  of the   partonic constituents of  the colliding  hadrons
  leads to the production  of intermediate subsystems (fireballs), which
  decay subsequently into hadrons.   The  weights of the  various  final
  states are derived from the corresponding phase-space factors modified
  by  empirical transition elements. The  results compare well with data
  at  energies    between   particle  production  thresholds   and   ISR
  energies. Special emphasis is put on correlation data, which offer the
  opportunity  to  shed  some  light on   the question  whether particle
  production proceeds via fireballs or strings.
%
} 
\maketitle
\section{Introduction}
\label{intro}

QCD is assumed to be the theory of the  strong interaction.  Soft hadron
production, however,   is  a nonperturbative   process which at  present
cannot be calculated  by  QCD.  Thus, the  understanding of  soft hadron
production  is  still based on   phenomenological  approaches like, {\it
e.g.}, the dual parton (DPM)~\cite{capella94}, the VENUS~\cite{werner93}
or  the PYTHIA-LUND~\cite{andersson83,andersson87,sjostrand87a,pythia94}
models.  Within these  approaches  hadrons are considered  as  composite
objects consisting  of partons  (quarks and  gluons) the  interaction of
which is   assumed to proceed   in two steps.  In   a first step excited
subsystems, usually called strings,  are produced which decay afterwards
into stable particles and resonances.

Ingredients of these models are structure functions, parton-parton cross
sections  and fragmentation functions.  The   structure functions of the
interacting   hadrons  contain   the   information about   the  momentum
distribution  of partons.   From  the  cross sections for  parton-parton
interactions the number  of strings produced  in a single  hadron-hadron
scattering can be   deduced, and  fragmentation  functions describe  the
decay of strings into hadrons.  These approaches make maximal use of the
information available  from lepton-hadron and lepton-lepton interactions
as  well as  from general properties   of the scattering  amplitude like
unitarity and analyticity.

On the  other hand  the first  attempts   to understand multiple  hadron
production in  hadronic   interactions    were  based    on  statistical
considerations and the   observation of excited intermediate  subsystems
called fireballs (FB) (see the reviews
\cite{feinberg71,dremin78,feinberg83}).     A  modern    version  of   a
thermodynamical    approach  can      be    found, {\it    e.g.},     in
ref. \cite{becattini97}.

The         Rossendorf            collision          (ROC)         model
\cite{muellerh95,muellerh90,muellerh91,muellerh92,muellerh95a,muellerh96,muellerh96a} is basically  a statistical approach in deriving the
relative contributions  of   the various final   channels by calculating
their statistical weights   from  the phase-space factors,  which   are,
however, strongly modified by empirical transition  matrix elements.  It
aims at  describing soft hadron  production in the energy region between
particle  production thresholds and ISR energies.   For this purpose the
basic ingredients of hadron production models are reformulated in such a
way that they are applicable  at low energies as well.  The ROC model is
built as a  minimal approach in the sense  that the number of parameters
is restricted to   the  minimum necessary to   well  reproduce the  main
features of  the  available  data.   Special emphasis   is   put on  the
consideration of short-range correlations, because the FB concept yields
a natural explanation of the observed phenomena.

The present paper is organized as follows.   In sect.~\ref{mo} the basic
features  of the ROC model are  explained.   Section~\ref{ex} contains a
comparison of selected  experimental data with   model calculations.  In
order to show differences  between string and  FB approaches with regard
to correlation  data the  ROC model  is  contrasted with the PYTHIA-LUND
model~\cite{pythia94}.  By means of a special version  of the ROC model,
where  FBs are degenerated into single  hadrons, the importance of FBs
for the  description of correlation   data is demonstrated.  Conclusions
are summarized in sect.~\ref{con}.

\section{The model}
\label{mo}
The basic idea of a statistical approach consists in the assumption that
the  probabilities   of formation   of  the  various  final  states  are
proportional to their statistical weights.  This idea was implemented by
Fermi~\cite{fermi50}   fifty years ago, but  his  model turned out to be
applicable only at relatively low energies.  At higher energies it is no
longer a  good approximation to assume  that the whole initial energy is
randomly distributed among   the final  particles. Particles with   high
transverse momentum, {\it  e.g.},    are produced with   extremely   low
probability indicating that the final states are dynamically linked with
the initial state.

\begin{figure}[tbp]
   \psfig{file=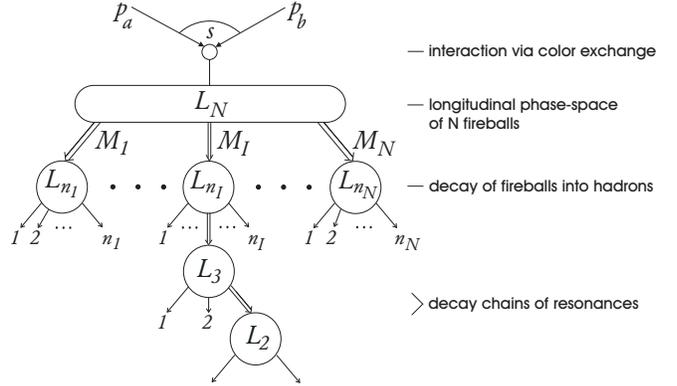,width=8.6cm}
   \caption[]{\label{ph}  Phase-space   decomposition of   a    two-step
   process.  $N$  FBs with masses $M_I (I=1  \dots N)$ are  produced in
   the interaction  of two hadrons $a$ and  $b$ with  four-momenta $p_a$
   and  $p_b$.  The FBs  decay in the second step,  where the $I$-th FB
   disintegrates into  $n_I$ primary  particles  with $\sum_{I=1}^N  n_I
   =n$.  Resonances  among the  primary particles decay  afterwards into
   stable hadrons.  A possible decay chain is shown.}
\end{figure}

In   contrast to  this early   attempt the  ROC  model  is based on  the
following  modified  statistical approach.   Instead  of calculating the
statistical  weights  from the  whole  phase-space the  dynamics  of the
interaction is  implemented in form of  empirical functions which either
suppress certain  regions    of the phase-space   or  impose  additional
non-statistical   weights.  We    define   a channel  $\alpha$ by    the
number~$n$, masses~$m_i$ and quantum numbers of the final particles. The
relative probability of populating a  channel $\alpha$ is calculated  as
the   product  of     the  Lorentz-invariant   phase-space factor   $\di
L_n(s;\alpha)$ with the  square  of an empirical  matrix  element $A^2$,
which describes the dynamics  of the interaction process.  Here, $s=p^2$
denotes   the square  of the  total  energy  with $p$    being the total
four-momentum.  The  phase-space factor is  defined as the integral over
the momenta of the final particles with energy and momentum conservation
taken into account,
\begin{equation} \label{phase-space1}
  \di L_n(s;\alpha) = \di L_n(s;m_1,\dots,m_n)
  =   \prod_{i=1}^n\frac{\di ^3p_i}{2e_i}   \, \delta^4
  (p-\sum_{i=1}^n p_i),
\end{equation}
where   the  four-momentum  of   the $i$-th    particle  is denoted   by
$p_i=(e_i,\vec{p}_i)$  with $p^2_i=m^2_i$.   For numerical  calculations
the   $\delta$~function   in  eq.~(\ref{phase-space1})  is   removed  by
introducing a   new  set  of $3n-4$    variables to replace    the  $3n$
three-momentum   components.    It is reasonable   to    choose a set of
variables, which   reflects   the underlying physical picture    of  the
interaction process.  In fig.~\ref{ph} the reaction
%
\[
   a + b \to FB_1 + \dots + FB_N \to h_1 + \dots + h_n
\]
%
between hadrons $a$ and $b$ resulting in the production of $n$ particles
is schematically depicted.  First, $2 \le N \le n$ intermediate particle
groups  called FBs are   produced, which  decay into  so-called  primary
particles.  The   primary particles define  the channels  for  which the
weights (\ref{prob1}) are calculated. Among  them are resonances,  which
decay subsequently into stable hadrons.
The phase-space factor corresponding to the diagram of fig.~\ref{ph} can
be calculated according to ref. \cite{byckling73}
\begin{eqnarray} \label{phase-space2}
  \di   L_n\,(s;\an)      &  =   &     \left[\prod_{I=1}^N \di     M_I^2
  L_{n_I}(M_I;\alpha_I)\right] \nonumber \\ & & \di L_N(s;M_1,\dots,M_N)
\end{eqnarray}
with     the     invariant  masses    of        the   FBs  equal     to
$M_I=\sqrt{P_I^2}=\sqrt{\left(\sum_{i=1}^{n_I}p_i\right)^2}$   and   the
final channel defined as the vector $\an = (\alpha_1,\dots,\alpha_N)$ of
the decay channels of the individual FBs. The probability of populating
the channel $\an$ is given by
\begin{equation} \label{prob1}
  \di W(s;\an) \propto \di L_n(s;\an) \, A^2 \;.
\end{equation}
Here,  the  square of  the  matrix element $A^2$  contains the dynamical
input and is split into factors
%
\[
  A^2=    A^2_{\mrm i} A^2_{\mrm   {qs}}   A^2_{\mrm {ex}} A^2_{\mrm  t}
  A^2_{\mrm l} A^2_{\mrm {st}}\,,
\]
%
which describe  the  interaction   $A^2_{\mrm   i}$ resulting  in    the
production of $N$~FBs, the production of  hadrons $A^2_{\mrm {qs}}$ via
the creation of quark-anti-quark ($q\bar{q}$) pairs , the invariant-mass
distribution of the FBs  $A^2_{\mrm ex}$, the transverse $A^2_{\mrm t}$
and longitudinal $A^2_{\mrm l}$ momentum  distribution of the FBs, and,
finally, some additional   factors $A^2_{\mrm {st}}$ necessary  for  the
calculation of the statistical   weights.  In the  following subsections
these factors will be discussed in more detail.
\subsection{The interaction $A^2_{\mrm i}$}
\label{mo-inter}
In models like   DPM   \cite{capella94} or VENUS   \cite{werner93}   the
colliding hadrons are  considered  as  extended and composite    objects
consisting of an indefinite number  of partons, the interaction of which
is assumed to proceed via  color exchange.   If,  {\it e.g.}, the  color
exchange between the valence quarks of two baryons takes place, then two
strings are produced. Each of them consists of the remaining diquark and
the valence quark removed from the other baryon.  Since more complicated
exchanges are possible and the collision may proceed at different impact
parameters  a varying number  of strings is produced.  The corresponding
probabilities  are derived from   Gribov-Regge theory \cite{gribov68} in
the limit   of high  energies in  combination  with the  use  of profile
functions    for     integrating     over   the   impact      parameter.
PYTHIA~\cite{pythia94} describes low-$p_{\mrm t}$ events on the basis of
the multiple    interaction model  of ref.    \cite{sjostrand87a}, which
extends  a high-$p_{\mrm  t}$ picture  down   into the low-$p_{\mrm  t}$
region  by  regularizing the   $p_{\mrm   t}$  scale.   The   number  of
(independent)   parton-parton collisions in  one   event depends on  the
impact parameter   and on the   assumed  matter distribution  inside the
interacting hadrons.

In the    ROC~model,  non-statistical weights    $A^2_{\mrm  i}(N)$  for
producing a  definite  number  $N$ of FBs   are  introduced.  This  is a
phenomenological  parameterization  of  the    contributions   from  the
different  color exchange diagrams and  of  the integral over the impact
parameter, which we apply also at low energies.   Such a dynamical input
is necessary, because the phase-space factor alone tends to overestimate
the number  of FBs  simply  due to the   fact that the number  of states
increases the more FBs are produced.  A thermodynamical approach using a
chemical potential as the only  parameter to regulate  the number of FBs
turned out to be not flexible  enough.  Therefore, the negative binomial
distribution is applied where we have two free parameters $v$ and $q$
\begin{equation} \label{NB}
   A^2_{\mrm i}(N;q,v)= {-v \choose N} (-q)^N (1-q)^{v}.
\end{equation}
In the calculations  we use  the mean $\bar{N}$  and  the  ratio of  the
variance  to the mean  $D = \sigma^2  /\bar{N}$ as parameters from which
$q=(D-1)/D$ and $v=\bar{N}/(D-1)$ follow.  By    this means, the   whole
complicated  interaction scenario    is  described  by  altogether   two
parameters in conjunction with the corresponding phase-space factors.
\subsection{Quark statistics $A^2_{\mrm {qs}}$}
\label{mo-quark}
The factor  $A^2_{\mrm  {qs}}$ stands   symbolically  for the  algorithm
applied to   sample the possible final  states.    At first  the valence
quarks of  the interacting hadrons are redistributed  among the FBs and
then the final   hadrons   are  produced   via  the  creation  of    new
$q\bar{q}$~pairs.    All   internal    quantum   numbers are   conserved
automatically by this procedure.

The  complicated details of color exchange  diagrams and string drawings
are  replaced  by  statistical   considerations.  For    describing  the
interaction process  the notion of quark  removal is  borrowed from ref.
\cite{werner93}.  It is assumed that the multiple interaction of the
partonic  constituents of the incoming  hadrons leads to the creation of
$N$ FBs. Color exchange between the  constituents results in the removal
of the involved  quarks and gluons from the  incoming hadrons.   Removed
quarks  are found in any of  the other FBs  with equal probability.  The
remaining partons  of the interacting hadrons form  the two leading FBs,
the  scattered partons  the    $N-2$  central  FBs.   In   a  next  step
$q\bar{q}$~pairs are produced  and  randomly (the $q$'s and  $\bar{q}$'s
independently)  distributed between the  FBs  such that each  FB becomes
color  neutral and contains the minimal  number of $q$'s and $\bar{q}$'s
necessary for  building at  least one hadron  (meson or  baryon).   This
procedure is the equivalent for the sum over the possible color exchange
diagrams  with the restriction that the  removal of two or three valence
quarks or of a single sea  quark is neglected.   Only the removal of one
valence quark with probability $W_{\mrm v}$ is considered. The remaining
probability $1-W_{\mrm v}$ is understood as gluon or quark-pair removal.
As equivalent to string fragmentation each FB  is then filled separately
with an arbitrary  number $n_{qI}$ of  additional $q\bar{q}$~pairs where
$n_{qI} \ge 0$.  Up, down, strange and charm quarks  are produced in the
ratios
\begin{equation} \label{uds}
	u:d:s:c=1:1:\lambda_{\mrm s}:\lambda_{\mrm c}
\end{equation}
with $\lambda_{\mrm s}$ and $\lambda_{\mrm c}$ being suppression factors
due to the heavier masses of the strange and charm quarks.  The creation
of top quarks can be neglected in the considered energy range.

The final hadrons are built up in each FB independently according to the
rules   of quark statistics   \cite{anisovich73}  by randomly  selecting
sequences of  $q$'s and $\bar{q}$'s.   A $q\bar{q}$ gives a meson, while
baryons or antibaryons are formed from $qqq$ or $\bar{q}\bar{q}\bar{q}$.
From a  given  sequence  of quarks  the  different  hadrons  are  formed
according  to the tables  of the particle  data group~\cite{PDG98}.  All
baryons marked in the tables with three or four  stars, the meson nonets
built  from u, d   and s  quarks   with angular  momenta  zero   and one
($^1S_0,^3S_1,^1P_1,^3P_0,^3P_1,^3P_2$) as  well as  all charmed hadrons
are taken into account.  An empirical probability distribution
\begin{equation} \label{probmass}
   W_{\mrm h}(m_{\mrm h}) \propto \exp{(-m_{\mrm h}/\Theta_{\mrm h})}
\end{equation}
with  an adjustable parameter~$\Theta_{\mrm h}$  is used to suppress the
formation  of the  heavier hadrons of   mass $m_{\mrm h}$.  During event
generation the current  masses of resonances  are sampled according to a
probability  distribution consisting of the   product of a  relativistic
Breit-Wigner distribution, the phase-space factor  of the decay products
and the  above  suppression  factor  $W_h(m_{\mrm  h})$.  The  decay  of
resonances into the various channels proceeds  either according to known
probabilities  or in  accordance with the    statistical weights of  the
possible final states in case of unknown decay probabilities.

The described algorithm together with the parameters $\lambda_{\mrm s}$,
$\lambda_{\mrm   c}$ and $\Theta_{\mrm h}$   is   the equivalent of  the
usually much larger number of parameters describing the fragmentation of
strings.

It should be stressed  that the ROC model  has  no parameter fixing  the
probability of diffractive processes.  A  diffractive process is usually
assumed to proceed via the exchange of a  Pomeron, a fictitious particle
which does not affect the quantum numbers of the involved particles.  In
the  string models  \cite{capella94,werner93}  diffractive scattering is
treated as  a special process whose  probability is determined by a free
parameter adapted  to data.  In  the ROC model diffractive scattering is
one  of the   possible   final channels,  because  there   is a  certain
probability that one or  even both leading  FBs  are identical  with the
initial  protons.   This happens if  the  valence quark  content  of the
considered FB remains  unchanged, if no additional $q\bar{q}$~pairs  are
produced,  and if a proton,   and not a   resonance,  is built from  the
available quarks $uud$ in the recombination phase.
\subsection{Mass distribution of FBs $A^2_{\mrm {ex}}$}
\label{mo-ex}
Until now we have explained how the  hadrons forming the final state are
sampled.  In a  next step the integral  over the invariant masses of the
FBs  $M_I$  [see     eq.~(\ref{phase-space2})] is   performed.     Again
phase-space  alone produces  too   large  FB masses  because   of  their
corresponding large numbers of states.  To restrict the invariant masses
the FBs are  assumed to be  characterized by a temperature  $\Theta$. As
matrix element squared the function
\begin{equation} \label{Aex}
  A^2_{\mrm ex} = \prod_{I=1}^N (M_I/\Theta)  K_1(M_I/\Theta)
\end{equation}
with the asymptotic behavior
\begin{eqnarray*} \label{AexI}
  (M_I/\Theta) K_1(M_I/\Theta)& {\longrightarrow \atop  M_I/\Theta
  \to     \infty}   &  \sqrt{M_I/\Theta}   \exp(-M_I/\Theta)      \\
  (M_I/\Theta) K_1(M_I/\Theta)& {\longrightarrow  \atop M_I/\Theta
  \to 0}& 1
\end{eqnarray*}
is used. The expression  $(M_I/\Theta) K_1(M_I/\Theta)$ is the kernel of
the    so-called   $K$-transformation   (see~\cite{byckling73})  used to
transform a micro-canonical  phase-space  distribution depending on  the
total energy $M_I$  of  the $I$-th FB  into  a canonical one,   which is
characterized   by a  temperature   $\Theta$.  In eq.~(\ref{Aex})  $K_1$
stands  for   the modified Bessel    function.  For increasing invariant
masses   $M_I$ the  function  $(M_I/   \Theta) K_1(M_I/\Theta)$ strongly
decreases, while the phase-space factor  $L_{n_I}(M_I)$ of the $I$-th FB
becomes  larger.   Their  product has a   maximum at  a  value  of $M_I$
determined by the   parameter  $\Theta$, which thus  fixes   the average
internal excitation energy of the $I$-th  FB.  Since FBs consist of only
a few particles, we are far  from the thermodynamical limit.  Therefore,
the momenta of the   hadrons are calculated from  the  decay of the  FBs
according to phase-space and not from macro-canonical distributions.
\subsection{Longitudinal phase-space of FBs $A^2_{\mrm t}$ 
and $A^2_{\mrm l}$}
\label{mo-tl}
FBs   are  proposed to emerge  from   the interaction  of partons, whose
momenta inside fast moving hadrons  are mainly longitudinal. Due to  the
finite size of hadrons and  the uncertainty principle a small transverse
component is present  too.  This is taken into  account by damping large
transverse momenta of the FBs using a linear exponential distribution
\begin{equation} \label{At}
  A^2_{\mrm t} = \prod_{I=1}^N \exp(-\gamma P_{\mrm {t,I}})
\end{equation}
with the mean $\bar{P}_{\mrm t}=2/\gamma$.   The parameter $\gamma$ used
here can  be directly compared with the  analogous parameter employed in
string models~\cite{capella94,werner93}.   There, the transverse momenta
of  the  partons, which form  the  ends of  the strings, are restricted,
here, we constrain the transverse momenta of the FBs directly.

The  longitudinal  momentum distribution of   the two  leading  FBs  is
weighted by
\begin{equation} \label{Al}
  A^2_{\mrm l} = (X_1 X_2)^\beta
\end{equation}
with the scaling variables
\[
  X_1=(E_1+P_{\mrm    z,1})/\sqrt{s}         \quad     \mbox{and}  \quad
  X_2=(E_2-P_{\mrm z,2})/\sqrt{s}\,.
\]
Here, it is assumed that FB~1 is the remnant of  the incoming hadron $a$
moving in the positive z-direction prior  to the interaction, while FB~2
stems from hadron $b$ moving in the  opposite direction.  On the average
the leading FBs carry the largest part of  the longitudinal momenta as a
consequence  of the weighting (\ref{Al}).  This  forces the other FBs to
have  accordingly less longitudinal momenta.   In this manner the factor
$A^2_{\mrm l}$   is the equivalent of   the structure functions  used in
refs. \cite{capella94,werner93}.

The method of calculating  the  longitudinal phase-space of $N$~FBs   is
taken from ref. \cite{carey78} with appropriate modifications due to the
presence of $A^2_{\mrm l}$.
\subsection{Statistics $A^2_{\mrm {st}}$}
\label{mo-st}
In conclusion, all factors still necessary for a correct counting of the
final states are collected in the term
\begin{eqnarray} \label{Ast}
  A^2_{\mrm  {st}}({\vec\alpha_N})  &  = & \left\{  \prod_{I=1}^N
  g(\alpha_I)      \left(\frac{V}{\left(2\pi\right)^3}  \right)
  ^{n_I-1}   \right.      \nonumber  \\ &     &   \left.  \left[
  \prod_{i=1}^{n_I}(2\sigma_i+1)2m_i
\right] \right\} \left(\frac{V}{\left(2\pi\right)^3}\right)^{N-1}.
\end{eqnarray}
It   contains the spin  degeneracy   factors $(2\sigma_i+1)$, the
volume $V$   in which the  particles  are produced with  $V=4 \pi R^3/3$
determined by the  radius parameter  $R$.  The quantity  $g(\alpha_{\mrm
I})=\left(\prod_\beta n_\beta!\right)^{-1}$ is the degeneracy factor for
groups of $n_\beta$ identical particles in the final state of the $I$-th
FB and prevents multiple counting of identical states.
\subsection{Differential cross section}
Summarizing the above considerations  we define the  number of states in
the decay channel $\alpha_I$ of one FB
\begin{eqnarray} \label{ZI}
  \di{\cal      Z}_I(\alpha_I)        &   =   &    g(\alpha_{\mrm    I})
  \left(\frac{V}{(2\pi)^3}\right)^{n_I-1}
  \left\{\prod_{i=1}^{n_I}(2\sigma_i+1)2m_i\right\} \nonumber \\ & & \di
  M_I \left(  \frac{M_I}{\Theta} \right)  K_1   \left(\frac{M_I}{\Theta}
  \right) \di L_{n_I}(M_I;\alpha_I)
\end{eqnarray}
and the analogous number for the set of FB states
\begin{eqnarray} \label{ZN}
  \di{\cal Z}_N(s)    & =  &   \left(  \frac{V}{(2\pi)^3}  \right)^{N-1}
  \left\{\prod_{I=1}^N   2M_I\exp(-\gamma   P_{\mrm{t,I}})     A^2_{\mrm
  i}(N)\right\}    \nonumber     \\      &   &(X_1    X_2)^\beta     \di
  L_N(s;M_1,\ldots,M_N)\,.
\end{eqnarray}
From  eqs.   (\ref{ZI}) and  (\ref{ZN})  a  compact expression   for the
probability of populating the channel $\an = (\alpha_1,\ldots,\alpha_N)$
[see eq.~(\ref{prob1})] can be derived
\begin{equation} \label{prob2}
  \di W(s;\an) = \left\{\prod_{I=1}^N \di {\cal Z}_I(\alpha_{\mrm
  I})\right\} \di {\cal Z}_N(s) \,.
\end{equation}
The corresponding cross section is written as
\begin{equation} \label{cross}
  \di \sigma(s) =  \sigma_{\mrm {in}}(s) \frac{  \di  W(s;\an) } {\sum_N
  \sum_{\an} \int \di W(s;\an)}\,,
\end{equation}
where   the   inelastic cross section   $\sigma_{\mrm   {in}}(s)$ of the
considered reaction serves  as normalization.  Any physical quantity  of
interest can  be  derived    from  eq.  (\ref{cross})  by   summing  the
contributions   from all channels   and  integrating over the unobserved
variables
\subsection{Adjusting parameters}
\label{mo-pa}
The parameters introduced in the previous subsections determine definite
features of    the  production process.    So   the radius parameter  is
responsible   for  the   multiplicity   $n$ of   final   particles.   In
eq.~(\ref{Ast}) the factor $V^{n-1}$  appears implying $R^{3(n-1)}$, and
increasing $R$ in  turn  means  that states  with  large number   $n$ of
produced particles get higher weights.  The temperature $\Theta$ and the
radius $R$ determine the  mean invariant mass  of the FBs,  $\Theta$ via
the excitation energy,   $R$ via the number  of   particles in the   FB.
Abundances of different hadron   species are fixed by  the probabilities
$\lambda_{\mrm s}$   and  $\lambda_{\mrm c}$  for  the  creation of  the
different  quark   flavors.    The   parameter   $\Theta_{\mrm h}$    in
eq.~(\ref{probmass}) influences  the selection of  hadrons consisting of
the same  valence quarks.  Finally, the  momenta  of the primary hadrons
are composed from the  superposition  of the  FB momenta, determined  by
$\beta$ and $\gamma$, with  the  relative velocities  of hadrons in  the
rest frames of the FBs, influenced by  $\Theta$.  There are correlations
between the parameters,  and   several parameter sets have   been  found
giving similar results.
\begin{figure}[t]
   \psfig{file=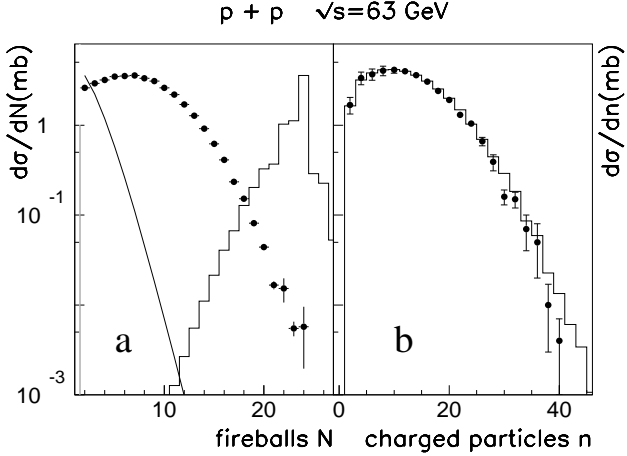,width=8.6cm}
   \caption[]{\label{AiF} The calculated FB~distribution~$\di \sigma/\di
   N$,      eq.  (\ref{dsdN}),     (dots  in    {\bf     a})    and  the
   measured~\cite{CDHW-br84b} and  calculated  multiplicity distribution
   $\di \sigma/\di  n$ of   charged particles  $n$  (dots and  histogram
   respectively     in    {\bf     b})    from   $pp$~interactions    at
   $\sqrt{s}=63\,\mrm{GeV}$.  In~{\bf  a} the dependence  of the  matrix
   element $A_{\mrm  i}^2(N)$~(line) and of  the FB distribution without
   $A_{\mrm i}^2(N)$   (histogram)  are depicted  separately  (these two
   curves are arbitrarily normalized).}
\end{figure}

In searching for  a  suited set of  parameters  we start at the  highest
energy $\sqrt{s}=63\,\mrm{GeV}$ with  a guess for  all parameters except
$\bar{N}$ and $D$, which are responsible for the  number of produced FBs
[see eq.(\ref{NB})].  These parameters are varied until the multiplicity
distribution of charged particles is reasonably reproduced.

In fig.~\ref{AiF} the influence of the factor $A_{\mrm  i}^2(N)$ on the
FB~distribution 
\begin{equation} \label{dsdN}
  \frac{\di \sigma(s)}{\di N}  = \sigma_{\mrm {in}}(s)  \frac{\sum_{\an}
  \int \di W(s;\an) } {\sum_N \sum_{\an} \int \di W(s;\an)}
\end{equation}
is demonstrated.  Without the factor $A_{\mrm i}^2(N)$ the cross section
for producing $N$  FBs (the  histogram  in fig.~\ref{AiF}a) reaches  its
maximum at   values   much too high  for   reproducing  the multiplicity
distribution of   charged  particles.    By  combining  the   increasing
phase-space factor with the decreasing  function $A_{\mrm i}^2(N)$  (the
line in fig.~\ref{AiF}a) a  FB distribution is produced which reproduces
the distribution of charged particles (fig.~\ref{AiF}b).  The similarity
of FB and particle multiplicity distribution is due  to the fact that on
the average each FB emits the same number of particles.

\begin{figure}[b]
   \psfig{file=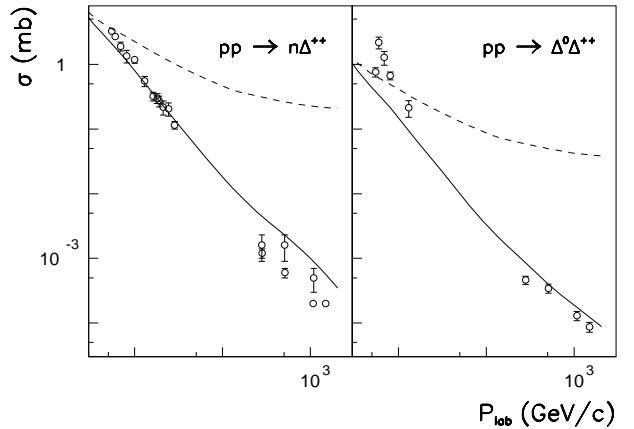,width=8.6cm}
   \caption[]{\label{AqF} Cross   section  of the reactions   $pp  \to n
   \Delta^{++}$  and $pp \to  \Delta^0\Delta^{++}$ as a  function of the
   laboratory momentum.  Experimental  data (dots) from \cite{landolt88}
   are compared with ROC model results (solid  lines). The dashed curves
   are  calculated with   the  probability  for valence  quark   removal
   $W_{\mrm v}=1$.}
\end{figure}
In order to describe data also at lower energies the parameters $W_{\mrm
v}$,  $\gamma$, $\beta$ and  $R$ were  made  energy dependent, while the
other ones  are kept   constant.  For the   probability $W_{\mrm  v}$ of
valence    quark  removal  the  energy     dependence follows from   the
consideration of the binary reactions $pp  \to n\Delta^{++}$ and $pp \to
\Delta^0\Delta^{++}$    for  which data  in   a  wide energy  region are
available (see fig.~\ref{AqF}).   In the present approach  contributions
to these  reactions  come alone from  events with  $N=2$ FBs. A  removed
quark from  one proton belongs, after  the interaction, to the other one
and vice versa. For the reaction to proceed an exchange of an $u$ with a
$d$~quark must happen according to
%
\[
   p+p \to uud+uud \to udd+uuu \to n+\Delta^{++}\;.
\]
The calculated cross sections become zero  for $W_{\mrm v}=0$, while for
$W_{\mrm v}=1$ the data are well reproduced at low, but overestimated by
orders of   magnitude   at high energies  (see    the dashed   curves in
fig.~\ref{AqF}). Therefore, the function
\begin{equation} \label{Wv}
   1-W_{\mrm v} = E_{\mrm k}/(W_1+E_{\mrm k})
\end{equation}
with  the excess  energy $E_{\mrm  k}=\sqrt{s}-2m_p$  and  one parameter
$W_1$ is  used to describe the  energy dependence.  It provides a smooth
transition  between the two  extremes, $W_{\mrm v}=1$  for excess energy
$E_{\mrm k}=0$ and $W_{\mrm  v} \to 0$  for  large excess  energies.  In
this way  the data  are reproduced quite  well  with the correspondingly
adjusted   parameter  $W_1$ (see    table~\ref{para}).   Such an  energy
dependence is in agreement with the  parton picture of hadrons.  Roughly
speaking the number of partons  increases with energy and, consequently,
the probability for a valence quark being involved in the color exchange
diminishes.
\begin{table}
   \caption{The parameter set used for the calculations. ME stands for
   the matrix element squared.}
   \label{para}
   \begin{tabular}{cllc}
   \hline\noalign{\smallskip}
   ME & Parameter & Parameter & Equation  \\
   \noalign{\smallskip}\hline\noalign{\smallskip}
   $A^2_{\mrm i}$    & $\bar{N}$=1.8 & D=1.4 & \ref{NB} \\
   $A^2_{\mrm {qs}}$ & $\lambda_{\mrm s}$=0.15 & $\lambda_{\mrm c}$=0.05
                     &\ref{uds}\\
                     & $\Theta_{\mrm h}=250\,\mrm{MeV}$ & 
                                 $W_1=3\,\mrm{GeV}$   
                     & \ref{probmass},\ref{Wv} \\
   $A^2_{\mrm {ex}}A^2_{\mrm {st}}$   & $\Theta=300\,\mrm{MeV}$ 
	&R=1.2$\,\dots\,$1.29 fm &\ref{Aex},\ref{Ast}\\
   $A^2_{\mrm t}$    & $\gamma_0=3.8\,\mrm{(GeV/c)^{-1}}$ 
                     & $\gamma_1=1\,\mrm{GeV}$ & \ref{At},\ref{eGam}\\
   $A^2_{\mrm l}$    & $\beta_0=2$ 
                     & $\beta_1=1\,\mrm{GeV}$  & \ref{Al},\ref{eBet}\\
   \noalign{\smallskip}\hline
   \end{tabular}
\end{table}

The energy  dependence of the parameters  $\gamma$ and $\beta$ [see eqs.
(\ref{At}), (\ref{Al})] cannot be determined in such a clean way.  Still
the best indication comes from the proton spectra at different energies.
In a diffractive event at least one of the leading FBs is identical with
the  corresponding initial  hadron.    The additional  folding with  the
momentum distribution  from the FB   decay is absent and the  transverse
momentum dependence of the diffractive peak (see figs. \ref{exDiPr} and
\ref{exDiBre} in sect.~\ref{exDi}) is  directly governed by $\gamma$ and
$\beta$.  A reasonable description at the different energies is achieved
using
\begin{equation} \label{eGam}
   \gamma = \gamma_0 \sqrt{s} / (\gamma_1+\sqrt{s})
\end{equation}
and
\begin{equation} \label{eBet}
   \beta  = \beta_0  \sqrt{s} / (\beta_1+\sqrt{s})\,.
\end{equation}

The parameter $R$ is then adjusted to reproduce the mean multiplicity of
charged   particles $\langle  n_{\mrm{ch}}\rangle$   at  the   different
energies.   Below incident  momenta  of $100\,\mrm{GeV/c}$  the constant
value  of      $R=1.29\,\mrm{fm}$ is used,    because   $\langle n_{\mrm
{ch}}\rangle$ ceases   to be  sensitive   to the value    of $R$.   With
increasing energies   the    value of $R$  decreases   smoothly  towards
$1.2\,\mrm{fm}$.  All ROC model  results  in this paper are   calculated
with the parameter set summarized in table~\ref{para}.
\section{Comparison with data}
\label{ex}
The  ROC~model is implemented as  a Monte-Carlo  generator which samples
complete events from which nearly all kinds of measurable quantities can
be  deduced and   compared   with experimental   results.   The  overall
agreement  between data and ROC calculations  is quite good in the whole
considered energy  range between about $\sqrt{s} \approx 2.2\,\mrm{GeV}$
and the    highest  ISR~energy   of  $\sqrt{s}=63\,\mrm{GeV}$.    In the
following a few selected data sets  concerning hadron multiplicities and
the  dependencies  of differential  cross  sections  on longitudinal and
transverse variables are  presented.   Then correlations  are  discussed
more  thoroughly, because   they are sensitive   to differences  between
string and FB models.
\subsection{Multiplicities}
\label{exMu}
\begin{figure}[t]
   \psfig{file=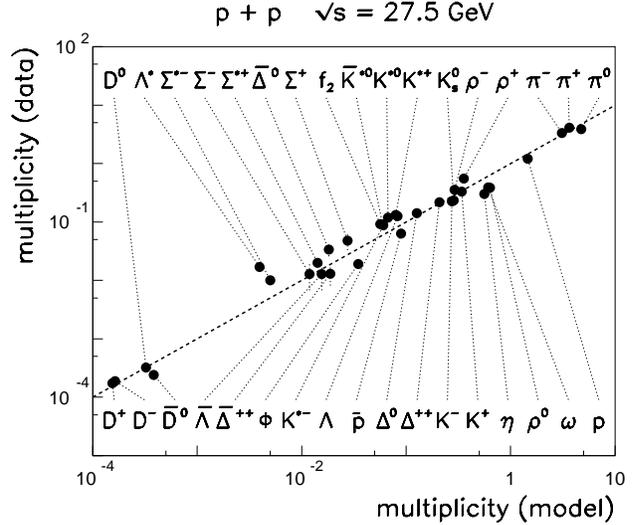,width=8.6cm}
   \caption[]{\label{exMuF1} Hadron  multiplicities  for $pp$~collisions
   at      $\sqrt{s}=27.5\,\mrm{GeV}$.     The    experimental   average
   multiplicities  \cite{landolt88,LEBC-EHS-ag91,kichimi79}  are plotted
   versus  the calculated ones.  The  dashed  line indicates coincidence
   between data and calculations.  Well reproduced data tend to lie near
   this line. $\Lambda^\ast$ stands for $\Lambda(1520)$.}
\end{figure}
Hadron abundances  are determined by  the strange  and charm suppression
factors $\lambda_{\mrm s}$ and $\lambda_{\mrm c}$, respectively, and the
hadron temperature $\Theta_{\mrm  h}$.    There is no parameter    which
governs the production of baryons directly.   Instead, the algorithm for
building hadrons from quarks is responsible  for baryon production.  The
more quarks are available in the FB considered, the  more probable it is
to select at  random  a sequence of  three  $q$'s (or $\bar{q}$'s)  from
which  a  baryon (or antibaryon)  can be  formed. Since   the FBs become
bigger with increasing temperature   $\Theta$ and radius $R$, these  are
the parameters, which determine how many baryons  are produced.  A rather
complete  data   set of hadron   multiplicities  for  $pp$~collisions at
$\sqrt{s}=27.5\,\mrm{GeV}$   is  compared   with  model calculations  in
fig.~\ref{exMuF1}.  The  agreement is  quite  impressive except for  the
$\Lambda(1520)$ where the deviation is  rather large.  An obvious reason
for this discrepancy has not been found.

Figure~\ref{exMuF1} should    be  compared with   the result    from the
thermodynamical    model     of     Becattini    (see     fig.~5      in
ref. \cite{becattini97}).   There, three parameters, temperature, volume
and suppression factor $\lambda_{\mrm s}$ are fitted, which have similar
meanings as  in  our   study, although   the parameter  values    differ
considerably from those used  here.  Nevertheless, the achieved accuracy
of  the   description is  comparable.   We    have,  contrary  to   ref.
\cite{becattini97}, charmed particles included in fig.~\ref{exMuF1}, and
the suppression factor $\lambda_{\mrm c}$ is adjusted to reproduce these
data points.
\begin{figure}[t]
   \psfig{file=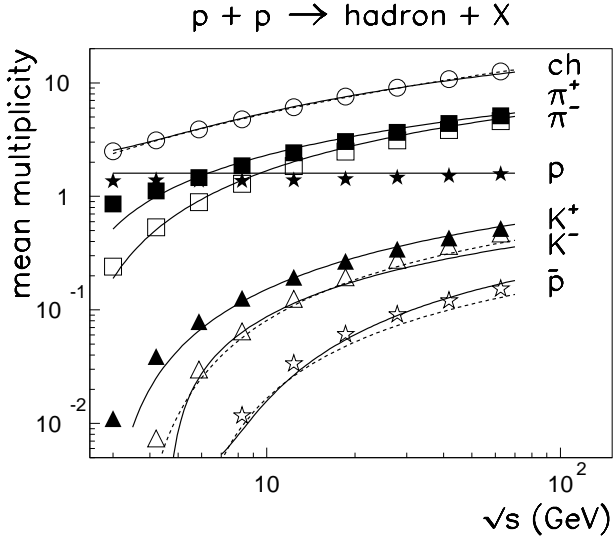,width=8.6cm}
   \caption[]{\label{exMuF2}    The   energy    dependence  of      mean
   multiplicities of charged particles  ($ch$), $\pi^+$, $\pi^-$, $K^+$,
   $K^-$, $p$  and $\bar{p}$ for  $pp$~scattering. Full and dotted lines
   are fits to the data~\cite{rossi75}, the symbols are ROC~results.}
\end{figure}

In  fig.~\ref{exMuF2} the energy  dependence  of mean  multiplicities of
different particle species is plotted.  The mean multiplicity of charged
particles   $\langle n_{\mrm {ch}}\rangle$   is  mainly affected by  the
temperature $\Theta$ and the radius $R$.  In order to reproduce $\langle
n_{\mrm {ch}}\rangle$ the  radius $R$ is adjusted at  each energy.   The
adapted values of $R$ change by less than 10\%  in the considered energy
region.  Since charged particles are mainly pions the energy dependences
of  the  number of charged particles   and of  pions   are similar.  The
slightly larger values of $\pi^+$  compared to $\pi^-$ arise from charge
conservation.  Strangeness suppression causes  a large gap between pions
and kaons.  It   should  be noted  that,  {\it  e.g.}, the VENUS   model
reproduces  this  gap (see fig.   10.2  of ref.  \cite{werner93}) with a
suppression $u:d:s=0.43:0.43:0.14$ differing by  a factor of nearly  two
from the    value  used  here,  namely  $u:d:s=1:1:0.15$.    A  possible
explanation  for  this   difference  might  be    the mass    factor  in
eq.~(\ref{Ast}).  Strange particles  have heavier masses what gives them
a  larger relative weight.   On  the other  hand, the phase-space factor
becomes smaller the heavier the produced particles are.  Hence, a direct
comparison of the suppression  factors  as used in string  fragmentation
and  in the present  approach seems to  be difficult.  The production of
antiprotons is suppressed by about two orders of magnitude, because only
FBs consisting of at least three  quark pairs have a certain probability
to  create a  baryon-antibaryon   pair.   The direct  influence   of the
phase-space factor becomes  especially important in the threshold region
where  the multiplicities are  small.   The different curves for  pions,
kaons and antiprotons are well reproduced due to the phase-space factor,
which strongly decreases with diminishing kinetic energy still available
after particle production.
\begin{figure}[t]
   \psfig{file=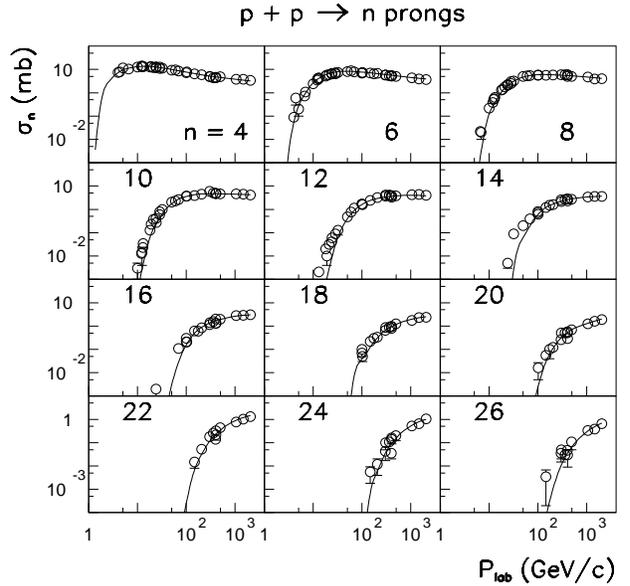,width=8.6cm}
   \caption[]{\label{exMuF3}   The topological cross sections $\sigma_n$
   for the  production  of $n$ charged  particles as  a function of  the
   laboratory  momentum.   Dots  are data~\cite{landolt88},  full  lines
   ROC~results.}
\end{figure}

After having reproduced the mean  multiplicities of charged particles by
adapting   the radius parameter we   consider  the energy dependence  of
topological cross  sections  in  fig.~\ref{exMuF3} without any   further
parameter adjustment. The typical behavior of  the cross sections with a
weak maximum  and subsequent  slow  decrease for   low and  a continuous
increase in the considered energy region for high multiplicities is well
reproduced by our model.
\subsection{Differential cross sections}
\label{exDi}
\begin{figure}[t]
   \psfig{file=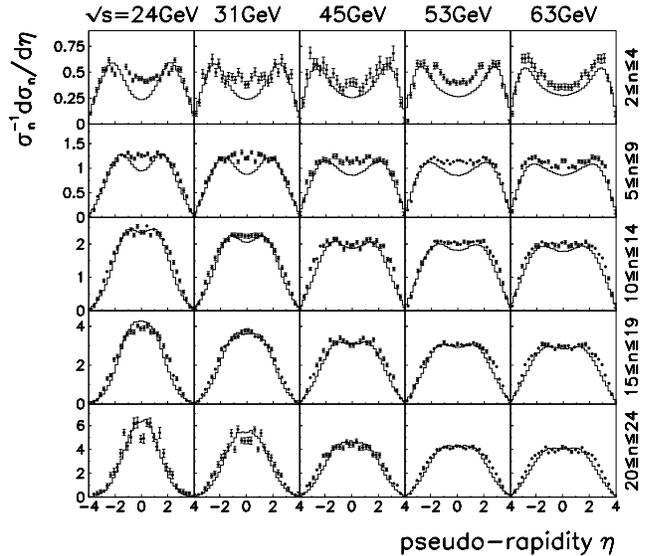,width=8.6cm}
   \caption[]{\label{exDiEta}  Normalized charged  particle densities in
   various intervals   of the multiplicity  $n$ indicated   on the right
   ordinate.   Dots are  data  \cite{thome77},  histograms represent ROC
   results.  Data  and calculations are corrected   for acceptance.  For
   the  data   $n$ is the   {\em observed}  multiplicity, while  for the
   calculations $n$ is  the  (true) multiplicity in  the pseudo-rapidity
   range $\eta \le 4$.}
\end{figure}
\begin{figure}[bt]
   \psfig{file=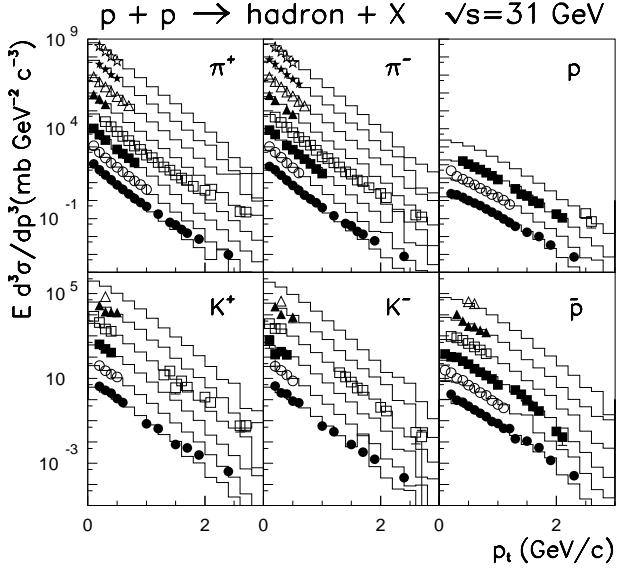,width=8.6cm}
   \caption[]{\label{exDiAl} Invariant cross  sections as  a function of
   the transverse momentum at cm.  rapidities $y$=0, 0.2, 0.4, 0.6, 0.8,
   1.0,  1.2  and    1.4   from  bottom  to   top.    Symbols are   data
   \cite{BS-alper75}, histograms ROC results. The spectra are multiplied
   by $10^0$ at y=0, $10^1$ at y=0.2\, $\dots$\, $10^7$ at $y$=1.4.}
\end{figure}
\begin{figure}[bt]
   \psfig{file=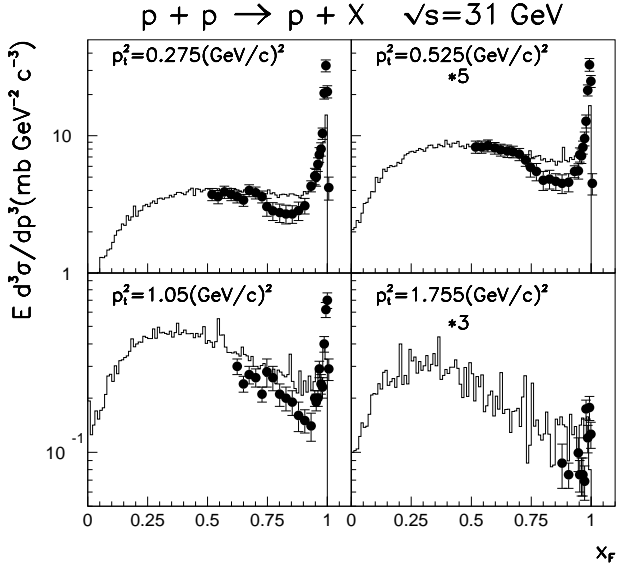,width=8.6cm}
   \caption[]{\label{exDiPr} Invariant differential  cross sections as a
   function  of Feynman's variable $x_F$  at  various values of $p_{\mrm
   t}^2$ as     indicated     in  the  figure.     Symbols     are  data
   \cite{CHLM-al73c}, histograms ROC results.  The  spectra on the right
   hand side are multiplied by factors of 5 and 3, respectively.}
\end{figure}
\begin{figure}[t]
   \psfig{file=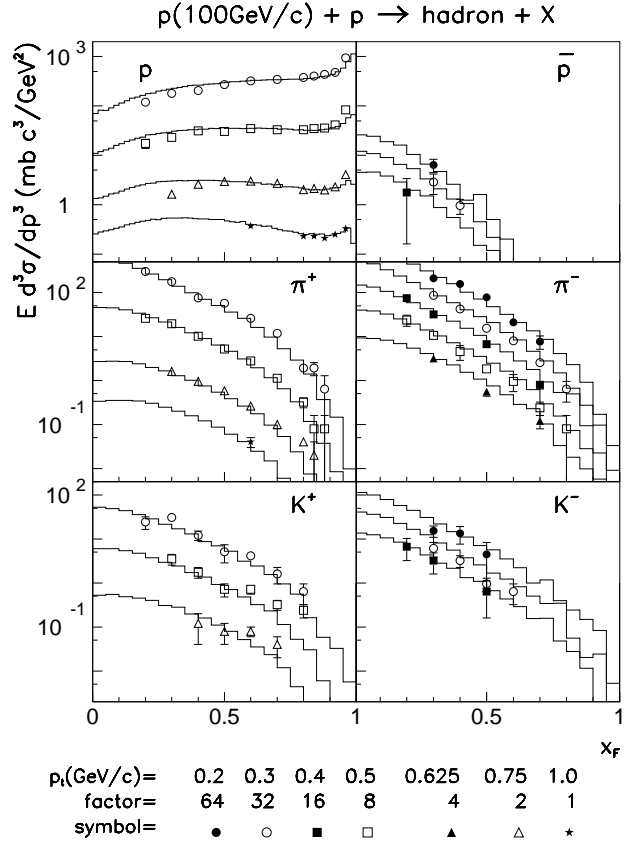,width=8.6cm}
   \caption[]{\label{exDiBre} Invariant  differential cross  section for
   the  production  of $p, \bar{p}, \pi^+,   \pi^-, K^+$ and $K^-$  as a
   function  of   Feynman's  variable $x_F$.    The  symbols  denote the
   data~\cite{brenner82} at   various values of the  transverse momentum
   $p_{\mrm t}$, histograms are ROC results.   Data and calculations are
   multiplied by the factors given in the legend.}
\end{figure}
\begin{figure}[t]
   \psfig{file=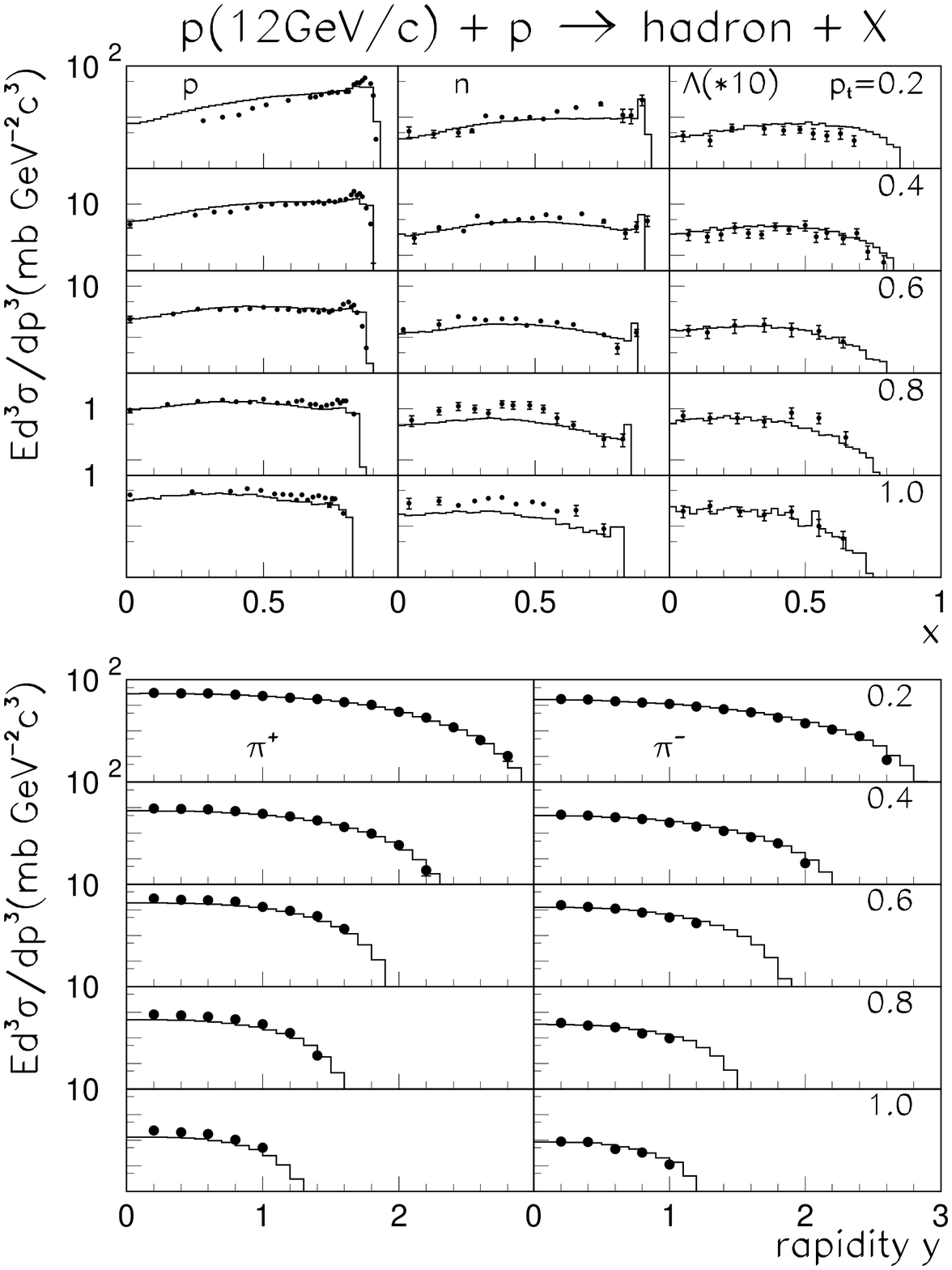,width=8.6cm}
   \caption[]{\label{exDiBlo} Invariant differential cross sections  for
   the production of protons, neutrons and  $\Lambda$'s as a function of
   $x=2p_l/\sqrt{s}$ (upper part) and of $\pi^+$ and $\pi^-$~mesons as a
   function of the  rapidity $y$ (lower  part).  The  symbols denote the
   data~\cite{blobel74,blobel78} at  values  of the transverse  momentum
   $p_{\mrm  t}$ between 0.2 and  $1.0\,\mrm{GeV/c}$, histograms are ROC
   results.}
\end{figure}
In  a next  step   the distribution   of   particles in  phase-space  is
considered.   We start  with  the charged particle  density  for various
intervals   of     the  {\em    observed}   multiplicity     shown    in
fig.~\ref{exDiEta}.   Charged  particles were measured~\cite{thome77} in
the pseudo-rapidity region    $|\eta|\le 4$.  In    the calculations the
geometrical acceptance of  the  apparatus  has been taken  into  account
according  to the curve  given in fig.~2  of  ref.  \cite{thome77}.  The
differences  in the multiplicities  {\em observed} in the experiment and
the multiplicities   in     the   calculations    are,  however,     not
corrected. Nevertheless, the  characteristic  features of the  data  are
excellently reproduced.   We see the two  bump structure at  low and the
shrinkage of the distributions with increasing multiplicities as well as
their broadening with increasing energy.   In the ROC calculations  this
behavior comes  from the  decay of the   leading FBs situated  near  the
projectile and the target   rapidity at low multiplicities,   while with
increasing multiplicity the  contributions from the increasing number of
FBs dominate and due to  momentum conservation the  mean rapidity of the
leading FBs change to lower values too.

The   same data were  described by   ref.  \cite{uhlig78}  with a simple
cluster model and by \cite{girija88} in the framework of the dual parton
model \cite{capella94}. Both attempts failed in reproducing the two-bump
structure  at low multiplicities,   obviously due  to  the absence  of a
diffractive component in these approaches.

In  fig.~\ref{exDiAl} transverse   momentum distributions   for  various
particles are compared  with  experimental data.   Both  data and  model
results show  roughly an exponential   behavior.  The calculated spectra
result from a convolution of the transverse momenta  of the FBs with the
internal momentum distributions of the primary particles  in the FBs and
the momentum distributions of   secondaries in which primary  resonances
decay.  At the highest momenta an  underestimation of the measured cross
section is  observed while the  overall agreement  is quite satisfactory
for all particle types.

Proton spectra as a function  of a longitudinal variable like  Feynman's
variable $x_F$   are of special    interest because  of  the  peak  from
diffractive scattering for $x_F \to 1$. As already mentioned there is no
special  parameter  which  forces this  channel    to  have a   definite
probability.  Therefore,  the good  overall  reproduction of  the proton
spectra in fig.~\ref{exDiPr} is quite remarkable, although the height of
the peak  is  underestimated by  the  calculations.  A  better agreement
might be obtained by giving the leading FBs a lower temperature than the
central ones    in analogy  to   the spectator-participant   picture  of
hadron-nucleus reactions.   This  ansatz  will  be  left to  forthcoming
considerations.

Also  at lower energies excellent  agreement  between data and ROC~model
calculations is achieved,  as demonstrated in fig.~\ref{exDiBre}.  There
the  dependence of  invariant  differential cross  sections on Feynman's
variable  $x_F$ for the    production of various  hadron  species   at a
laboratory momentum of $100\,\mrm{GeV/c}$ is depicted.  While the proton
spectra show  pronounced diffractive  peaks for $x_F  \to  1$, the cross
sections for the other particle types  strongly decrease at large $x_F$.
The   squared  longitudinal   matrix  element  $A^2_{\mrm l}$   with the
parameter $\beta$ in eq.(\ref{Al})  governs this behavior.  It keeps the
leading FBs  at large $|x_F|$ and the  particles from the central FBs at
small $|x_F|$.    The smaller $\beta$   is chosen  the  larger the cross
sections become for mesons and $\bar{p}$'s for $x_F \to 1$.
\begin{figure}[t]
   \psfig{file=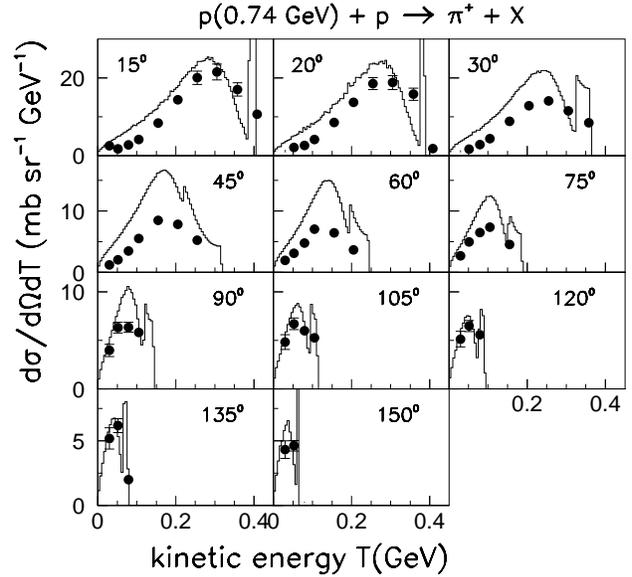,width=8.6cm}
   \caption[]{\label{exDiCo} Differential cross section as a function of
   the  kinetic energy   for $\pi^+$   production  at laboratory  angles
   between $15^\circ$  and $150^\circ$.  Dots are data~\cite{cochran72},
   histograms ROC results.}
\end{figure}

In stepping  down  the energy  scale let  us give  a further example  in
fig.~\ref{exDiBlo}.  Shown are  the invariant  cross sections for  three
types of baryons  as a function   of $x=2p_l/\sqrt{s}$ ($p_l$  being the
longitudinal  momentum) and for   charged pions as    a function of  the
rapidity .  The main features of the data  are well reproduced, although
again the diffractive peaks in the proton spectra are underestimated. In
case of  the neutron spectra there are  deviations  especially at higher
values of  $p_{\mrm t}$,  while   the $\Lambda$~spectra  are quite  well
described.  Also the rapidity  plateau at small  and its shortening with
increasing transverse momenta in the pion spectra is well reproduced.

Finally, we verify the applicability of the ROC  model at energies below
the  production threshold  for strange particles  and  show the  kinetic
energy  spectra    of   $\pi^+$~mesons  at   an     incidence energy  of
$740\,\mrm{MeV}$ in fig.~\ref{exDiCo}.  The  peak at the high-energy end
of the calculated spectra comes from the binary reaction $pp\to d\pi^+$.
Due  to  insufficient energy resolution  this  peak  is not seen  in the
measured spectra.  The data are well  reproduced in forward and backward
direction  while deviations become  noticeable  at sidewards angles.  At
low energy  the parameter  values discussed   in sect.~\ref{mo}  are  no
longer important.  Instead,  the details of  the  treatment of resonance
decays play an important  role.   Most of the $\pi^+$~meson   production
proceeds   via the  creation  of $\Delta$  resonances.   The spectra are
therefore strongly influenced by  the mass distributions of the  decayed
resonances.  In the present version of the  ROC model the current masses
of all types of resonances are sampled by  using a constant width of the
Breit-Wigner  distribution.  This might be the   reason for the observed
deviations.

\subsection{Correlations}
\label{exCo}
The data considered  so far can  obviously  be quite well  reproduced by
both string and FB models.  In this subsection the question is discussed
whether rapidity  correlations are more  sensitive with respect to their
interpretation in terms of strings or FBs.  The existence of short-range
correlations  is experimentally well   established and many  papers deal
with the various aspects of correlations (see, {\it e.g.},
\cite{basetto71,hayot74,morel74,berger75,meunier75,benecke78}).

When  looking   for rapidity    correlations  one  usually    defines  a
single-particle
\[
\rho_1(y)=\sigma_{\mrm {in}}^{-1} \di\sigma/ \di y
\]
and a two-particle rapidity density
\[
\rho_2(y_1,y_2)=\sigma_{\mrm {in}}^{-1} \di^2 \sigma/ \di y_1 \di y_2.
\]
The latter is proportional to the probability of finding one particle at
$y_1$  and a  second one  at  $y_2$. In order  to  see whether the joint
production of a   pair of  particles  at  $(y_1,y_2)$   differs from  an
independent production of the two particles the two-particle correlation
function
\begin{equation} \label{C12}
	C(y_1,y_2) = \rho_2(y_1,y_2) - \rho_1(y_1) \rho_1(y_2)
\end{equation}
is introduced as the difference between the two-particle density and the
product  of   two single-particle   densities. Non-vanishing  values  of
$C(y_1,y_2)$ indicate the presence of correlations.
\begin{figure}[t]
   \psfig{file=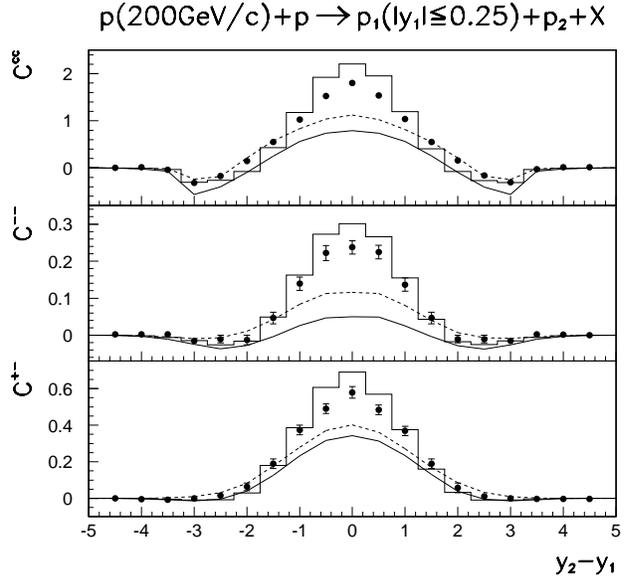,width=8.6cm}
   \caption[]{\label{exCo200} The    inclusive  two-particle correlation
   function $C(y_1,y_2)$ from $200\, \mrm  {GeV/c}$ $pp$ interactions as
   a  function of $y_2-y_1$  with  fixed $|y_1| \le  0.25$ for different
   charge  states of the  two observed particles  indicated by $C^{cc}$,
   $C^{--}$  and $C^{+-}$.  Dots  are data~\cite{whitmore76}, histograms
   ROC~model, full lines PYTHIA and dashed lines ROCS results.}
\end{figure}
\begin{figure}[t]
   \psfig{file=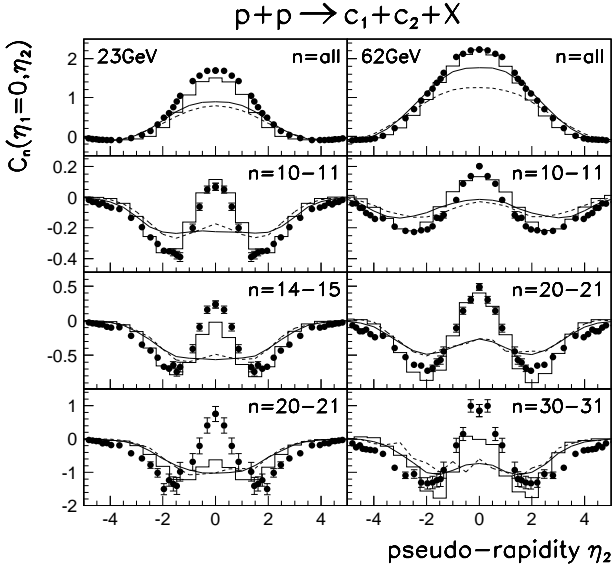,width=8.6cm}
   \caption[]{\label{exCoISR-0}  Inclusive   (n=all)  and semi-inclusive
   two-particle correlations $C_n(\eta_1,\eta_2)$  vs. $\eta_2$ at fixed
   $\eta_1=0$ for the {\em observed} multiplicities $n$ indicated in the
   figure    from $pp$~interactions  at   $\sqrt{s}=23\,\mrm{GeV}$   and
   $62\,\mrm{GeV}$.  At $23\,\mrm{GeV}$ the corresponding mean values of
   the {\em true} multiplicities are 9.1,  14.0 and 21.7, and the curves
   are  calculated for    multiplicities   of   8-10,  14 and     18-24,
   respectively, while at $62\,\mrm{GeV}$ the  mean values are 8.1, 18.8
   and 30.8,   and the curves are calculated   for multiplicities  of 8,
   18-20 and  28-34, respectively.   Dots  are  data~\cite{amendolia76},
   histograms  ROC~model,  full lines  PYTHIA   and  dashed  lines  ROCS
   results.}
\end{figure}
\begin{figure}[t]
   \psfig{file=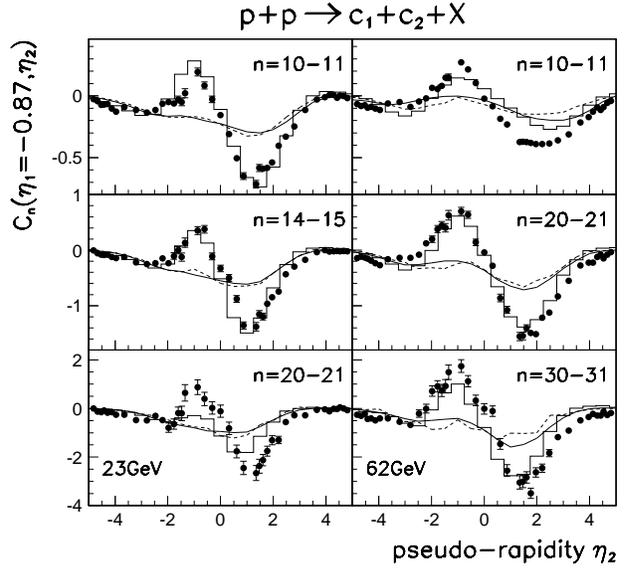,width=8.6cm}
   \caption[]{\label{exCoISR-0.87} The same as fig.~\ref{exCoISR-0}, but
   for semi-inclusive two-particle correlations $C_n(\eta_1,\eta_2)$ vs.
   $\eta_2$ at fixed $\eta_1=-0.87$.}
\end{figure}

In fig.~\ref{exCo200}    the   measured  and   calculated   two-particle
correlation   functions for  different charge  combinations   of the two
observed  particles are compared.   The pronounced peaks in the measured
spectra are satisfactorily    reproduced by the ROC~model   calculations
(histograms).   However, this fact   alone is  not  yet a  proof  of the
existence of FBs for at least two  reasons.  First, the measurements are
inclusive ones and mixing  of  events with different  multiplicities can
cause strong   correlations     as,   {\it  e.g.},    pointed   out   in
refs. \cite{morel74,eggert75a,SPS-UA5-an88}.   Second, the presence   of
resonances among the emitted particles  also tends to group the observed
particles into clusters.  That  means, the observed  correlation spectra
contain  always a superposition of   effects  from resonances and  other
possible short-range phenomena.

In order to  see whether resonance  production  alone can reproduce  the
observed  correlation patterns we carry out  calculations with a special
version of the  ROC~model (abbreviated by  ROCS in  the following).  All
FBs  are compelled to degenerate into  single hadrons by restricting the
quark  content of the FBs to  be either $qqq$ or $\bar{q}\bar{q}\bar{q}$
or $q\bar{q}$.  In this way all correlations implied  by the presence of
FBs are excluded.  An additional  parameter $W_{\mrm B}$ is necessary in
this  case fixing the probability  of baryon  creation relative to meson
production,  because the algorithm  for the combined building of baryons
and mesons  from  quarks  (see sect.~\ref{mo-quark}) is   not applicable
here.   The value of $W_{\mrm   B}=0.15$ adjusted to reproduce $\bar{p}$
production is similar to  the probability of  diquark creation in string
models,   where, {\it     e.g.},      the default  values    used     in
VENUS~\cite{werner93}  and   PYTHIA~\cite{pythia94} are  0.12  and 0.10,
respectively.  The ROCS version has  been proven to reasonably reproduce
most  of the data  discussed  in the previous  sections with  readjusted
parameters  of $A^2_{\mrm i}$    [see eq.~(\ref{NB})] and an   increased
radius parameter $R$.  The ROCS results for the correlation are shown in
fig.~\ref{exCo200} by the dashed lines, which underestimate the measured
values remarkably. We  consider this result  as a direct verification of
the presence of short-range phenomena beyond resonance production.

As a further proof of such additional short-range phenomena calculations
with the  string model PYTHIA~\cite{pythia94}  version 6.115 are carried
out using the multiple  interaction approach of ref. \cite{sjostrand87a}
with varying  impact   parameter.   The mean   multiplicity  of  charged
particles  is reproduced by adjusting  the regularization scale $p_{\perp
0}$ (PARP(82)) of the transverse momentum spectrum.  A Gaussian is taken
for  the matter  distribution of  the interacting protons,  what gives a
reasonable  reproduction  of the  multiplicity distribution  of  charged
particles at $\sqrt{s}=63\,\mrm{GeV}$. All  other parameters are kept at
their  default      values.  The     results,   the   full      lines in
fig.~\ref{exCo200}, underestimate the  data too.  This finding  confirms
again  that resonance production  alone is  insufficient  for  a correct
reproduction of  correlation data. In  addition, the  similarity of ROCS
and PYTHIA  curves shows that the  fragmentation of  multiple strings in
PYTHIA can be  quite well imitated  by creating hadrons  in longitudinal
phase-space.

Correlations  caused   by    the  mixing  of   events    with  different
multiplicities  can  be excluded  by  fixing  the multiplicities of  the
considered events.  This     type   of measurements  will     be  called
semi-inclusive in the following.   For the correlation function the same
formula (\ref{C12})  holds  with the  single and two-particle  densities
taken from events having a definite multiplicity.  As an example results
of   Amendolia {\it   et  al.}  \cite{amendolia76}    are presented   in
fig.~\ref{exCoISR-0} for the reaction $p +  p \to c_1 +  c_2 + X$ at two
energies  with $c_1$, $c_2$ and  $X$  standing for two charged particles
and anything, respectively.   The inclusive two-particle correlation (on
top of the figure) is compared with semi-inclusive correlations at three
narrow intervals of the {\em observed}  multiplicities.  The latter data
sets, however, contain events   whose {\em true} multiplicities  span  a
much  wider interval (see fig.~5 in  ref.  \cite{amendolia76}).  Lacking
the   exact knowledge  of  the    detector  response  we have   selected
multiplicity  intervals  around the   mean  values  of   the {\em  true}
multiplicities~\cite{amendolia76} in the calculations.  Figure
\ref{exCoISR-0} shows  that the inclusive  correlation  function is much
broader than the semi-inclusive ones,  a feature  of  the data which  is
well reproduced  by the ROC  calculations.  The semi-inclusive data with
their distinctly smaller peaks and and  pronounced dips of both sides of
the peaks are well described  too.  Alone the peaks  in the data for the
highest  multiplicity  especially  at    $\sqrt{s}=23\,\mrm{GeV}$    are
underestimated.   The  PYTHIA and   ROCS curves   clearly  underestimate
inclusive as  well as   semi-inclusive data.   Again   they are in  good
agreement with each other.

\begin{figure}[t]
   \psfig{file=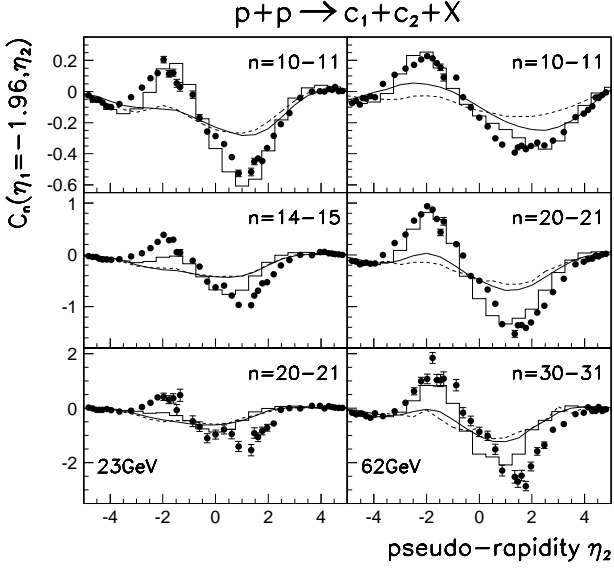,width=8.6cm}
   \caption[]{\label{exCoISR-1.96} The same as fig.~\ref{exCoISR-0}, but
   for   semi-inclusive  two-particle  correlations $C_n(\eta_1,\eta_2)$
   vs. $\eta_2$ at fixed $\eta_1=-1.96$.}
\end{figure}

In  fig.~\ref{exCoISR-0.87}  the correlation   function with one charged
particle   fixed  outside midrapidity is   given.    The  data are  well
reproduced   by the  ROC~model,   except for   the highest  multiplicity
interval at $\sqrt{s}=23\,\mrm{GeV}$.  These deviations become larger if
the distance of the fixed  particle from  midrapidity is increased  (see
fig.~\ref{exCoISR-1.96}).  The ROCS  version underestimates also  in the
non-symmetric   cases all considered   correlations  and agrees with the
PYTHIA results with the exception of the lowest multiplicity interval at
$\sqrt{s}=62\,\mrm  {GeV}$.  There, the PYTHIA curve  is   closer to the
experiment than the ROCS result, but still  far from a good reproduction
of the data.

\begin{figure}[t]
   \psfig{file=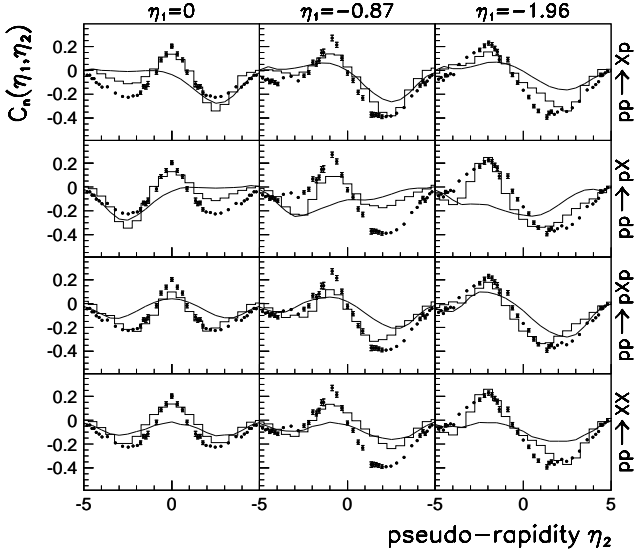,width=8.6cm}
   \caption[]{\label{exCoSub}  Semi-inclusive two-particle  correlations
   $C_n(\eta_1,\eta_2)$  vs.  $\eta_2$   at   several fixed  values   of
   $\eta_1$ indicated    in  the figure  for the   multiplicity interval
   $n=10-11$  at $\sqrt{s}=62\,\mrm{GeV}$.  The  data~\cite{amendolia76}
   (dots)  are compared with    ROC~model (histograms) and PYTHIA  (full
   lines)  results for  single- ($pp \to  Xp$  and $pp \to pX$), double-
   ($pp \to pXp$) and nondiffractive ($pp \to XX$) processes.}
\end{figure}

To better understand the results shown  in figs.   \ref{exCoISR-0}   -
\ref{exCoISR-1.96} the contributions from different subprocesses
are depicted in   fig.~\ref{exCoSub} separately.  In PYTHIA  diffractive
and nondiffractive processes are treated in different ways, while in the
ROC model   a diffractive process is  simply  one of  the possible final
channels without any special assumptions.  Consequently, the ROC results
for  all subprocesses exhibit the typical   correlation pattern.  On the
other  hand,  the  PYTHIA results   show  a large variety of  completely
different correlation functions    in   dependence on   the   considered
subprocess and on   the pseudo-rapidity   interval  where one   of   the
particles is   fixed.  In   the  first  row  of  fig.~\ref{exCoSub}  the
diffractive  excitation of  the projectile is   considered.  We find the
target  proton  after the interaction   at negative  rapidities  and the
excited projectile as a string in case of PYTHIA and as  a number of FBs
in case  of ROC calculations mainly at  positive  rapidities.  Values of
the correlation function around  zero near the target rapidity  indicate
that  the emission of  the target proton is   weakly correlated with the
emission of the charged particles from the excited subsystem(s) (see the
symmetric case $\eta_1=0$).
For   the diffractive  excitation of   the  target (the  second row) the
unchanged target proton  is    at positive, the  excited   subsystems at
negative  rapidities.   The  whole picture  is  simply  reversed  in the
symmetric  case ($\eta_1=0$),    while  we get from  PYTHIA   completely
different  pictures     for    the non-symmetric      measurements  with
$\eta_1=-0.87$  and  $\eta_1=-1.96$.    The  PYTHIA  result   for double
diffraction (the  third row) is of  special  interest.  Both interaction
partners remain unchanged and the excited subsystem is mainly at central
rapidities.   The central  string  acts like  a   cluster or FB and  the
results from  ROC  and PYTHIA are   similar.  The  contribution  of this
special process is responsible  for the somewhat better  reproduction of
the   data   by PYTHIA   in the   multiplicity   interval   $n=10-11$ at
$\sqrt{s}=62\,\mrm{GeV}$ in   fig.  \ref{exCoISR-1.96}.  With increasing
multiplicities the diffractive   contributions  become smaller and   the
result is dominated by the nondiffractive  component.  In this case (the
fourth row) we have  in PYTHIA the typical  multi-string picture  with a
clear  underestimation of  the measured correlation.   Altogether we see
that even  at  fixed  multiplicities the contributions   from  different
processes may result in    distinct  diffraction patterns in a     model
dependent way.

\begin{figure}[t]
   \psfig{file=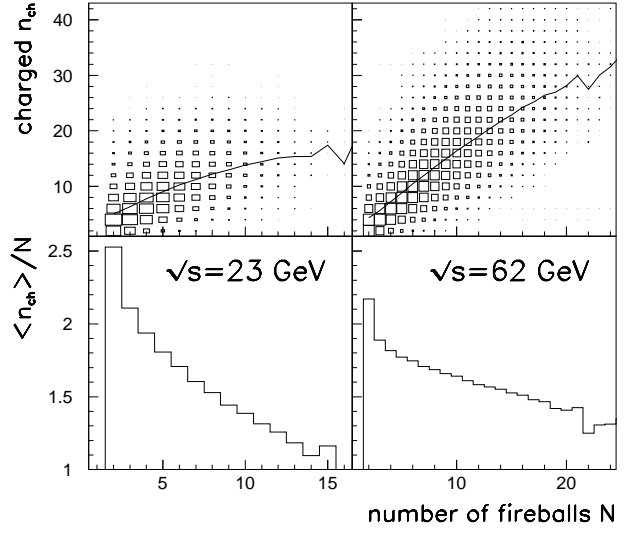,width=8.6cm}
   \caption[]{\label{exCoROC} Number  of   charged   particles  $n_{\mrm
   {ch}}$ versus number of  FBs $N$ in ROC  model calculations for $p+p$
   at  $\sqrt{s}=23\,\mrm{GeV}$ and  $62\,\mrm{GeV}$.   The size  of the
   boxes is proportional to the calculated cross section.  The full line
   in the  upper part  represents the  mean  value of charged  particles
   $\langle n_{\mrm {ch}}\rangle$ and the line  in the lower part the FB
   size $\langle n_{\mrm {ch}}\rangle/N$ as a function of N.}
\end{figure}
In spite of  the results presented here  it should be  mentioned that in
the literature examples for the  description of short-range correlations
by string models can be found.  So the dual parton model
\cite{capella94}, which has a  multi-string structure similar to PYTHIA,
reproduces  correlation data  either under  certain  assumptions for the
fragmentation functions  (see,  {\it e.g.}, \cite{pajares83,chudakov93})
or as the result  of the decay of minijets  and of soft strings with sea
quarks  at their  ends~\cite{bopp91}.   Such  strings emit hadrons  into
limited rapidity regions like clusters.

In  order   to clarify   the  origin  of   the  deviations  between  ROC
calculations   and     data     at     high     multiplicities      (see
fig.~\ref{exCoISR-1.96}) we show the correlation  between the number  of
FBs and the number of charged particles in fig.~\ref{exCoROC}.  First of
all we observe large  fluctuations regarding the FB  size, where FB size
is understood  as the mean number  of charged particles originating in a
FB.  At  $\sqrt{s}=62\,\mrm{GeV}$  it is possible  that,  {\it e.g.}, 18
charged particles may be produced in events with  numbers of FBs ranging
from 2 up to 20.  The mean number of  charged particles (the line in the
upper part of fig.~\ref{exCoROC}) flattens out  for large numbers of FBs
especially at  $\sqrt{s}=23\,\mrm{GeV}$.   That means that,  although we
use constant parameters $\Theta$ and $R$, the FB  size (the histogram in
the  lower part  of fig.~\ref{exCoROC}) becomes  smaller with increasing
number of  FBs due to the phase-space  factor, and this  influence is at
$\sqrt{s}=23\,\mrm{GeV}$ much stronger than at $62\,\mrm{GeV}$.  Here we
see the origin for the discrepancy with the data at high multiplicities.
This  is confirmed by a series  of calculations with increased parameter
values of $R$  (and accordingly adapted $A^2_{\mrm  i}$, eq.~(\ref{NB}),
to keep  the  mean number of  charged  particles constant).  Due to  the
larger $R$ the size of  the FBs is  increased and a good reproduction of
the  data at high multiplicities  is possible while the low-multiplicity
data are  overestimated.  That  means, the  present  version of  the ROC
model is too simple with regard to the  FB size.  The volume $V$ derived
from the parameter $R$ is considered as a measure for the overlap region
of the two  colliding hadrons  which  defines the spatial region   where
particle production takes place. Therefore, it  is understandable that a
value   of $V$   independent of   the  impact   parameter is  surely  an
oversimplification.  Since the number of produced FBs is correlated with
the   impact parameter (see sect.   \ref{mo-inter}),  an increase of $R$
with increasing number  of FBs would  be quite a  natural improvement of
the  model.  A  dependence  of the FB  temperature  $\Theta$ and  of the
momentum distribution  of FBs described by $\beta$   and $\gamma$ on the
number of  FBs  cannot be excluded  as well.   A  central  collision may
produce hotter FBs with a more isotropic phase-space distribution than a
peripheral one.
\begin{figure}[t]
   \psfig{file=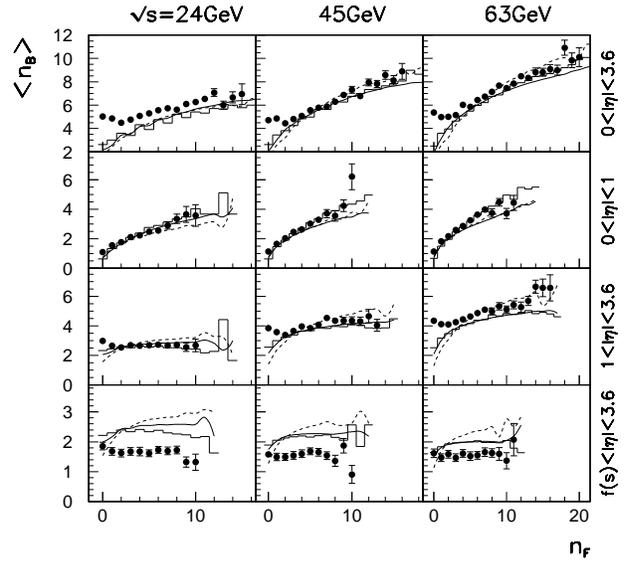,width=8.6cm}
   \caption[]{\label{exCoUh}  Mean  number of charged  particles emitted
   into  the backward hemisphere $\langle  n_B\rangle$  as a function of
   the number of  charged particles emitted  into the forward hemisphere
   $n_F$ for symmetric intervals of the pseudo-rapidity $\eta$ indicated
   in the figure,  where $f(s)=0.5 \cdot \ln(s \cdot \mrm{GeV}^{-2})-2$.
   Dots are data~\cite{uhlig78}, histograms ROC~model, full lines PYTHIA
   and dashed lines ROCS results.}
\end{figure}
Such  a  view  is supported    by an  analysis  \cite{FTC-E735-al95}  of
long-range correlations between  charged   particles emitted into    the
forward    and    backward  hemispheres   at     much   higher  energies
($\sqrt{s}=0.3-1.8\,\mrm{TeV}$).   The E735  collaboration  came to  the
conclusion, that the cluster size may increase as a function not only of
$\sqrt{s}$  but  also of  the   particle multiplicity.

An analysis  of long-rang correlations  in  the energy region considered
here has been  carried out by Uhlig {\it  et al.} \cite{uhlig78}.  Their
results  are shown  in fig.~\ref{exCoUh}.   The mean   number of charged
particles emitted into the backward hemisphere is  plotted as a function
of the number of forward  emitted particles.  The  slope of these curves
is  a measure  for the  strength  of  the correlation between  particles
ejected into the various regions of the hemispheres.  Although there are
some  systematic   deviations   at     low  multiplicities    and     at
$\sqrt{s}=24\,\mrm{GeV}$ the overall trend of the data is well described
by   the calculations.  We  see  the strongest  correlation for adjacent
regions in pseudo-rapidity $\eta$ while the strength is smaller if there
is a gap between the considered $\eta$~intervals. The emission is nearly
independent for the regions   in  the lowest row  of  fig.~\ref{exCoUh},
which are outside  the  central rapidity  plateau.  The  authors of ref.
\cite{uhlig78}  explain  their   results  with the  correlation  between
clusters  consisting of about three   particles (neutral ones included).
In fig.~\ref{exCoUh} we see a striking  agreement between the results of
ROC, ROCS and PHYTHIA calculations. Obviously, both FB as well as string
models are able to reproduce the observed long-range correlations in the
considered energy region.
\section{Conclusions}
\label{con}
We have presented  the empirical ROC  model for soft hadron  production.
It is based on the parton  picture of hadrons  as well as on statistical
and thermodynamical  considerations.    Experimental results  from  $pp$
interactions in the energy region between particle production thresholds
and  ISR energies  can  be  well described  with  a  moderate  number of
parameters which are either constant or have a smooth energy dependence.
A comparison of the ROC fireball model  and the PYTHIA string model with
regard   to  short-range rapidity  correlations   seems to favor the ROC
model, although the  description of short-range correlations is possible
in string models too.  Thus we do not have a clear answer to the problem
of  hadron  production    via  strings   or  fireballs.   A   systematic
consideration of all available  data especially those at higher energies
is necessary to    come to a decision  between   the possible scenarios:
strings, fireballs or something in between.

\begin{acknowledgement}
  The author would like to thank B.~K\"ampfer and especially H.-W.~Barz
  for valuable   discussions and    the    careful  reading   of     the
  manuscript. The  work is supported in part  by BMBF grants 06DR920 and
  06DR828/I.
\end{acknowledgement}

\bibliographystyle{EPJ}
\bibliography{hadMod,hmbib,hmhelp,ISR,Monte_Carlo,hh_data,High_Energy,DPM}
%
%
%

\end{document}